\documentclass[floatfix,11pt,nofootinbib]{revtex4}

\usepackage{amssymb,amsmath}
\usepackage{graphicx,float}
\usepackage{indentfirst}
\usepackage{slashed}
\begin{document}
	
	\title{Interesting examples of violation of the classical equivalence principle  but not of the weak one}

	\author{Antonio Accioly}
	
	\address{Coordena\c{c}\~{a}o de Cosmologia, Astrof\'{i}sica e Intera\c{c}\~{o}es Fundamentais (COSMO),  Centro Brasileiro de Pesquisas F\'{i}sicas (CBPF), Rua Dr. Xavier Sigaud 150, Urca, 22290-180,  Rio de Janeiro, RJ, Brazil}

	\author{Wallace Herdy}
	\address{Coordena\c{c}\~{a}o de Cosmologia, Astrof\'{i}sica e Intera\c{c}\~{o}es Fundamentais (COSMO),  Centro Brasileiro de Pesquisas F\'{i}sicas (CBPF), Rua Dr. Xavier Sigaud 150, Urca, 22290-180,  Rio de Janeiro, RJ, Brazil}
	%\email{wallacew@cbpf.br}

	\begin{abstract}
		The equivalence principle (EP), as well as Schiff's conjecture, are discussed {\it en passant}, and the connection between the EP and quantum mechanics is then briefly analyzed. Two semiclassical violations  of the classical equivalence principle (CEP)   but not of the weak one (WEP), i.e. Greenberger gravitational Bohr atom and the tree-level scattering of different quantum particles by an external weak higher-order gravitational field, are thoroughly investigated afterwards. Next, two quantum examples  of systems that agree with the WEP but not with the CEP, namely COW experiment and free fall in  a constant gravitational field  of a massive object described by its wave-function $\Psi$, are discussed in detail.  Keeping in mind that among the four examples focused on this work only  COW experiment is based on an experimental test, some important  details  related to it,   are presented as well.

	\end{abstract}

	%Uncomment for PACS numbers title message

	% Keywords required only for MST, PB, PMB, PM, JOA, JOB? 
	%\vspace{2pc}
	%\noindent{\it Keywords}: Article preparation, IOP journals
	% Uncomment for Submitted to journal title message
	%\submitto{\JPA}
	% Comment out if separate title page not required
	\maketitle

	\section{Introduction}
	The equivalence principle (EP) is intrinsically connected to the history of gravitation theory and has played an important role in its development. Newton regarded this principle as such a cornerstone of mechanics that he devoted the opening paragraph of the {\it Principia } to it. 
	
	Let us then discus, in passing, some important aspects related to  the EP in the framework of both Newton and Einstein gravity.

	The classical equivalence principle (CEP) of Newtonian theory (universality of free fall, or equality of inertial and gravitational masses) has a  {\it nonlocal }  character.		
	As far as Einstein gravity is concerned two EP are generally contemplated: the weak equivalence principle (WEP)  and the Einstein one (EEP). The WEP asserts that {\it locally} we cannot distinguish between inertial and gravitational fields through `falling body experiments'. Since the WEP, as well as the CEP, are   locally identical, the difficult, at the first sight, of differentiating them in an easy way increases. Consequently, some researchers are led to the common misconception that they coincide even nonlocally (see for instance \cite{1,2,3,4,5}). EEP, on the other hand, embodies   { \it WEP}, { \it local Lorentz invariance } --- the outcome of any local non-gravitational experiment is independent of the velocity of the freely-falling reference frame in which it is performed --- and { \it local position invariance} --- the outcome of any local non-gravitational experiment is independent of where and when it is performed \cite{6}.	
	EEP may be considered in the broadest sense of the term as the heart and soul of gravity theory. It would not be an exaggeration  to say that if the EEP holds, then  gravitation must necessarily be a `curved spacetime' phenomenon; in other words, the effects of gravity must be equivalent to the effects of living in a curved spacetime \cite{6}. 	
	Around 1960, Schiff conjectured that    { \it any complete and self consistent theory of gravity that obeys the WEP must also, unavoidable, obey the  EEP} \cite{7}. This surmise is known as Schiff's conjecture. According to it   the validity of the WEP alone should   guarantee the validity of the local Lorentz and position invariance, and thus of the EEP. However, a rigorous proof of Schiff's conjecture is improbable. In fact,  some special  counterexamples  are available in the literature \cite{8,9, 10, 11}. Nevertheless, there are some powerful arguments of `plausibility',  such as the assumption of energy conservation \cite {12} and the $TH \epsilon \mu$   formalism \cite{13}, among others, that  can be formulated.

	A natural question must now be posed: what is the connection between the   EP and quantum mechanics? As is well known, quantum tests of the EP are radically different from the classical ones  because classical and quantum descriptions of motion are fundamentally unlike. In particular, the universality of free fall (UFF) possesses a clear  significance in the classical  context. Now, how both UFF and WEP are to be understood in quantum mechanics is a much more subtle point. It is generally implicitly assumed that quantum  mechanics is valid in the  freely falling frame associated with classical test bodies. Nonetheless, an unavoidable problem regarding  quantum objects is the  existence of half integer spins, which have no classical counterpart. For integer spin particles, the EP can be  accounted for by a minimal coupling principle (see subsections A.1 and A.3 of Appendix A); while the procedure to couple a spin  $1/2$ field to gravity is much more complex and requires the use of a spinorial representation of the Lorentz group (see subsection A.2 of Appendix A).

	On the other hand, the most cited scientific experiment  claimed  to support the idea  that, at least in some cases, quantum mechanics and the WEP can be reconciled, is   COW experiment \cite{2}.
	Although this test,  as we shall prove, is in accord with the WEP, it is in disagreement with the CEP. Another example  of  a possible quantum mechanical  violation of the CEP but not of the WEP  is provided by  analyzing   free fall in a constant gravitational field of a massive object described by its wave-function $\Psi$.

	At the semiclassical level an interesting  event in which  the  CEP is also supposed to be  violated  but not the WEP is the tree level deflection of different quantum particles by an external weak higher-order gravitational field.	  We recall beforehand  that in Einstein theory  the scattering of any particle  by an external weak gravitational field is nondispersive  which, of course, is in agreement with the WEP. In other words,  the deflection angle of all massive  particles will  be exactly equal. The same is valid for the massless particles. Obviously, the deflection angle will be different whether the particle is massive or massless. A crucial question must then be posed: why to study  at the tree level the bending of  quantum particles in the framework of higher-derivative gravity? It is not difficult to answer this question. Higher-derivative gravity is the only  model  that is  known   to be renormalizable along its  matter couplings up to now \cite{14}. Nonetheless, since  this system is renormalizable, it is compulsorily nonunitary \cite{15,16}. We call attention to the fact that the breaking  down of  unitarity is  indeed a serious problem. Fortunately, we shall only deal with the linearized version of higher-derivative gravity, which is stable \cite{17}. The reason why it does not explode  is because the ghost cannot accelerate owing to energy conservation. Another way of seeing this is by analyzing the free-wave solutions. We remark that this model is not in disagreement with the result found by Sotiriou and Faraoni \cite{18}. In fact, despite containing a massive spin-2 ghost, as asserted by these authors, the alluded ghost cannot cause trouble \cite{19}. Another probable   example at the tree level of violation of the CEP but not of  the WEP is provided by Greenberger gravitational Bohr atom \cite{1}.	

	Our main goal here is to  explicitly show that in all situations described above, the WEP is not violated but the CEP is.

	The article is organized as follows.

	In Section 2 we study the following semiclassical examples:
	
	\begin{itemize}
		
		\item
		Greenberger gravitational Bohr atom.
		\item
		Tree-level scattering of different quantum particles by an external weak higher-order gravitational field.
		
	\end{itemize}

	After a careful  investigation of both models, we came to the conclusion  that  they do not violate at all the WEP but are not in accord with the CEP. As far as the second example   is concerned, it is worthy of note  that the resulting deflection angles  are  dependent on both spin and energy.  In addition  the well  known deflection angles (related to both massive and massless particles)   predicted by
	general relativity are recovered  through a suitable limit process.

	In Section 3  we analyze two  quantum examples:  COW experiment and  free fall in a constant gravitational field  of a massive object described in quantum mechanics by the wave-function $\Psi$. Again, these systems are in accord with the WEP but not with the CEP.

	Our comments are presented in Section 4.

	The lengthy calculations concerning the computation of  unpolarized cross sections  for the scattering of different quantum particles by an external weak higher-order gravitational field  are put in  Appendix A.

	We use natural units throughout and our Minkowski  metric is diag(1, -1, -1 ,-1).

	\section{Two examples of semiclassical violation   of the  CEP  but not  of the WEP } 
	
	We analyze in the following two examples  of semiclassical violation  of the CEP but not of the WEP  in a gravitational field.

	\subsection{Greenberger gravitational Bohr atom}  
	As  far as we know, Greenberger \cite{1} was the first  to foresee the existence of mass-dependent interference effects related to a particle bound
	in an external gravitational field.
	
	Here we are  particularly interested in analyzing Greenberger gravitational Bohr atom, which  from the classical point of view consists of a small mass $m$ bound to a very much 
	larger mass $M$ by the potential $V(r)=- \frac{GMm}{r}$, in the limit where  all
	recoil effects may be neglected. If we restrict  ourselves to circular orbits, we arrive at the conclusion that classically $\omega^2 r^3=GM $ (see figure 1).
	
	From this point on Greenberger  applied the same postulate proposed by Bohr:
	
	`The particles move in orbits restricted by the requirement that the angular momentum be an integer multiple of $\hbar$'.  Therefore, according to this postulate  for  
	circular orbits of radius $r$   the possible values of $r$ are restricted 
	by $L= mr^2 \omega =n$ \footnote{$\hbar =1$ since we are employing natural units. }, so that

	\begin{figure}[h!]
		\centering
		\includegraphics[scale=0.5]{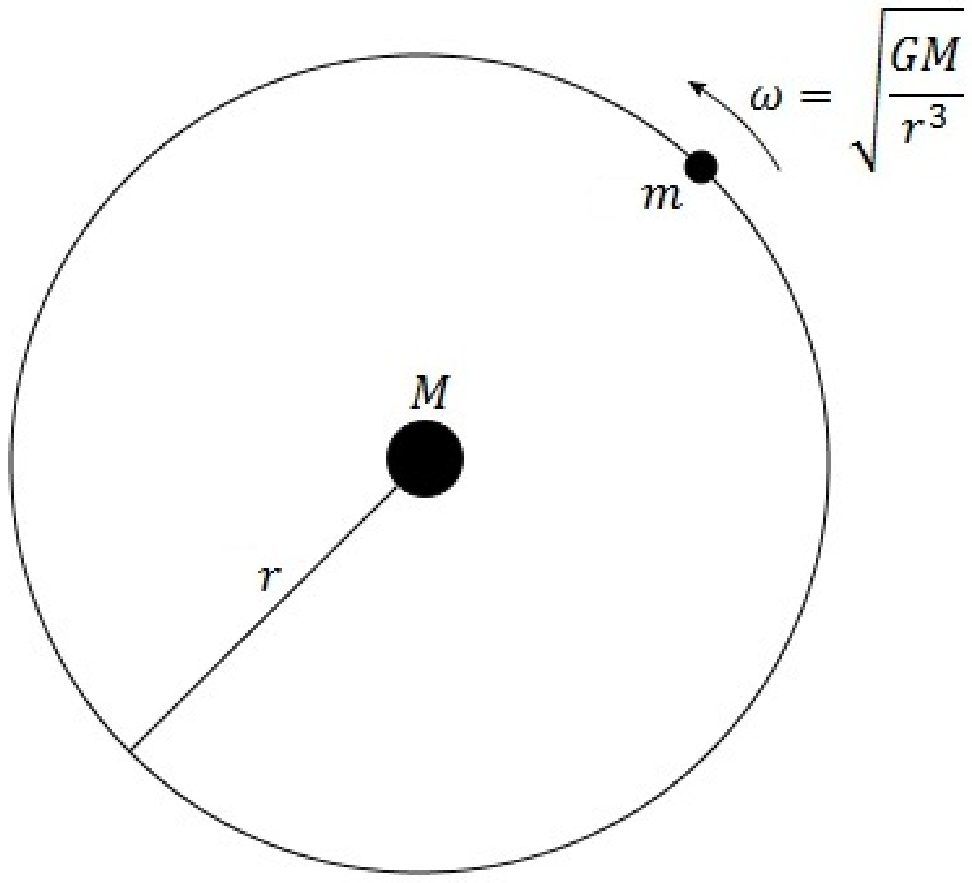}
		\caption{Greenberger  classical gravitational Bohr atom with circular orbit.}
		\label{fig}
	\end{figure}

	\begin{eqnarray}
	\omega_n= \frac{(GM)^2 m^3}{n^2}, \; \; r_n=\frac{n^2 }{GM m^2},\; \; E_n= \frac{(GM)^2 m^3}{2n^2}.
	\end{eqnarray}
	From the equations above, we see that lowest Bohr radius varies as $\frac{1}{m^2}$, and the orbital frequency  as $m^3$. As a consequence, it would  be trivial to tell the mass of the orbiting particle merely by observing its radius.    This result, of course, is in contradiction with what is expected from Newtonian gravity and the CEP. Nonetheless, there is no conflict between this result and the WEP. In fact, the WEP, as we have already mentioned, is a pure local statement, while Greenberger gravitational Bohr atom is an object  extended in space. Note however that the gravitational Bohr atom is not  a fully quantum system but  only a semiquantum or semiclassical one, exactly as it happens with the original Bohr's atom model, where according to the  aforementioned postulate the orbiting object has a well definite trajectory  and in addition there is the extra {\it ad hoc} assumption of quantization of the angular momentum.   In a fully quantum mechanical treatment, a probability of  presence is obtained via the wave-function, the `uncertainty principle' expressing the link between the width of the mentioned wave-function in both the direct and  reciprocal spaces.

	\subsection{Tree-level deflection of different quantum particles by an external weak higher-order gravitational field }

	The action  for higher-order gravity can be written as

	\begin{equation}
	I=\int \sqrt{-g} \Big[ \frac{2}{\kappa^2}R + \frac{\alpha}{2}R^2 + \frac{\beta}{2}R^2_{\mu \nu}\Big] - I_{\mathrm{M}},
	\end{equation}
	
	\noindent where $\kappa^2= 32 \pi G$, with $G$ being Newton's constant,  $\alpha$ and $\beta$ are free dimensionless coefficients, and $I_{\mathrm{M}}$ 
	is the action for matter.

	The field equations concerning the action  above are

	$$\frac{2}{\kappa^2}G_{\mu \nu} + \frac{\beta }{2}\Big[ -\frac{1}{2}g_{\mu \nu}R^2_{\rho \lambda} + \nabla_\mu \nabla_\nu R+ 2R_{\mu \rho \lambda \nu}R^{\rho \lambda}$$  $$- \frac{1}{2}g_{\mu \nu}\Box R - \Box R_{\mu \nu} \Big] + \frac{\alpha}{2}\Big[ -\frac{1}{2}g_{\mu \nu} R^2  + 2R R_{\mu \nu}$$ $$+ 2 \nabla_\mu \nabla_\nu R - 2 g_{\mu \nu} \Box R \Big]  + \frac{1}{2} \Theta_{\mu \nu} = 0, $$
	
	\noindent where $\Theta_{\mu \nu}$ is the energy-momentum tensor.
	
	From the above equation  we promptly obtain  its linear approximation doing exactly    as in Einstein's theory, i.e. we write
	
	\begin{equation}
	g_{\mu \nu}= \eta_{\mu \nu} + \kappa h_{\mu \nu},
	\end{equation}
	\noindent  and then linearize the equation at hand via (3), which results  in the following

	$$\Big( 1- \frac{\beta \kappa^2}{4}\Box \Big) \Big[ -\frac{1}{2} \Box h_{\mu \nu}  + \frac{1}{6 \kappa} R^{(\mathrm{lin})} \eta_{\mu \nu} \Big] + \frac{1}{2}(\Gamma_{\mu, \nu}$$ $$+ \Gamma_{\nu, \mu})= \frac{\kappa}{4} \Big( T_{\mu \nu} - \frac{1}{3}T \eta_{\mu \nu} \Big),$$

	\noindent where $$ R^{(\mathrm{lin})}= \frac{\kappa}{2}\Box h - \kappa {\gamma^{\mu \nu}}_{, \mu \nu},$$ $$  \gamma_{ \mu \nu} \equiv h_{\mu \nu} - \frac{1}{2}\eta_{\mu \nu} h,$$
	$$\Gamma_\mu \equiv \Big(1 - \frac{\beta \kappa^2}{4} \Box \Big) {\gamma_{\mu \nu}}^{ ,\nu} - \Big( \alpha + \frac{\beta}{2} \Big)\frac{\kappa}{2} {R^{(\mathrm{lin})}}_{, \mu}.$$
	
	\noindent  Note that indices are raised (lowered) using $\eta^{\mu \nu}$ ($\eta_{\mu \nu}$). Here $T_{\mu \nu}$ is the energy-momentum tensor of special relativity.

	It can be shown that it is always possible to choose  a coordinate system such that the gauge conditions, $
	\Gamma_\mu=0$, on the linearized metric hold. Assuming that these conditions are satisfied, it is straightforward to show that the general solution of the linearized  field equations is given by \cite{20,21}
	
	\begin{equation}
	h_{\mu \nu}= h^{({\mathrm{E}})}_{\mu \nu}- \phi \eta_{\mu \nu} + \psi_{\mu \nu},
	\end{equation}
	
	\noindent where $h^{({\mathrm{E}})}_{\mu \nu}$ is the solution of  linearized Einstein's equations in the de Donder gauge, i.e., 
	
	\begin{equation}
	\Box h^{({\mathrm{E}})}_{\mu \nu}= \frac{\kappa}{2} \Big[ \frac{T \eta_{\mu \nu}}{2} - T_{\mu \nu} \Big], \;\;\; \gamma^{{({\mathrm{E}}) \;,\nu}}_{\mu \nu} =0,  \nonumber
	\end{equation}
	
	$$\gamma^{{({\mathrm{E}})}}_{\mu \nu} \equiv h^{({\mathrm{E}})}_{\mu \nu} - \frac{1}{2}\eta_{\mu \nu} h^{({\mathrm {E}})},$$
	
	\noindent while $\phi$ and $\psi_{\mu \nu}$ satisfy, respectively, the equations
	
	$$\Big(\Box + m^2_0  \Big) \phi =\frac{\kappa T}{12},$$ $$\Big(
	\Box + m^2_2 \Big) \psi_{\mu \nu}=\frac{\kappa}{2} \Big[T_{\mu \nu} - \frac{1}{3}T \eta_{\mu \nu }\Big], \; \Box \psi= {\psi_{\mu \nu}}^{, \mu \nu}.$$
	
	\noindent It is worthy of note that in  this very special gauge the equations for $\psi_{\mu \nu}, \; \phi, \;$ and $h^{({\mathrm{E}})}_{\mu \nu} $ are totally decoupled. As a result, the general solution to the
	linearized  field equations reduces to an algebraic sum of the solutions of the  equations  concerning the three mentioned fields.
	
	Solving the preceding equations   for  a pointlike particle of mass $M$ located at ${\bf r=0}$ and having, as a consequence, an energy momentum tensor $T_{\mu \nu}= M\eta_{\mu 0} \eta_{\nu 0}\delta^3({\bf r})$, we find

	\begin{equation}
	h_{\mu \nu}(r)= h^{(\mathrm{E})}_{\mu \nu}(r) + h^{(\mathrm{R^2})}_{\mu \nu}(r) + h^{\mathrm{(R_{\mu \nu}^2)}}_{\mu \nu}(r) ,
	\end{equation}
	
	\noindent with
	
	$$h^{(\mathrm{E})}_{\mu \nu}(r)= \frac{M\kappa}{16 \pi} \Big[ \frac{\eta_{\mu \nu}}{r} - \frac{2\eta_{\mu 0} \eta_{\nu 0}}{r}\Big],$$ 
	
	$$h^{(\mathrm{R^2})}_{\mu \nu}(r)= \frac{M\kappa}{16 \pi} \Big[ -\frac{1}{3}\frac{e^{-m_0 r}}{r} \eta_{\mu \nu} \Big],$$ 
	
	$$h^{\mathrm{(R_{\mu \nu}^2)}}_{\mu \nu}(r)=\frac{M\kappa}{16 \pi} \Big[ - \frac{2}{3}\frac{ e^{-m_2 r }}{r}\eta_{\mu \nu} + 2 \frac{e^{- m_2 r}}{ r}\eta_{\mu 0} \eta_{\nu 0}\Big].$$

	\noindent Note that for $m_0, \; m_2 \rightarrow \infty$, the above solution reproduces the solution of linearized Einstein field equations in the de Donder gauge, as it should.
	
	On the other hand,  the momentum space gravitational field, namely  $h^{\lambda \rho}_{\mathrm{ext}}({\bf{k}})$, is defined by
	
	\begin{equation}
	h^{\lambda \rho}_{\mathrm{ext}}({\bf{k}})= \int{d^3{\bf{r}} e^{-i{\bf{k}}\cdot {\bf{r}}}h^{\lambda \rho}_{\mathrm{ext}}({\bf{r}})}.
	\end{equation}
	
	\noindent Thence,
	
	\begin{eqnarray}
	h^{\lambda \rho}_{\mathrm{ext}}({\bf{k}})= h^{{\mathrm{(E)}} \lambda \rho}_{\mathrm{ext}}({\bf{k}}) + h^{ {\mathrm{(R^2_{\mu \nu})}}\lambda \rho}_{\mathrm{ext}}({\bf{k}}) + h^{{\mathrm{(R^2)}}\lambda \rho}_{\mathrm{ext}}({\bf{k}}),
	\end{eqnarray}
	
	\noindent with
	
	$$h^{{\mathrm{(E)}} \mu \nu}_{\mathrm{ext}}({\bf{k}}) = \frac{\kappa M}{4{\bf{k}}^2}\eta^{\mu \nu} - \frac{\kappa M}{2} \frac{ \eta^{\mu 0} \eta^{\nu 0}}{\bf{\bf{k}}^2},$$
	
	$$h^{{\mathrm{(R^2_{\alpha \beta})}} \mu \nu}_{\mathrm{ext}}({\bf{k}})= - \frac{\kappa M}{6} \frac{\eta^{\mu \nu}}{{\bf{k}}^2 + m^2_2} +\frac{\kappa M}{2}\frac{\eta^{\mu 0} \eta ^{\nu 0}}{{\bf{k}}^2 + m^2_2}, $$
	
	$$h^{{\mathrm{(R^2)}} \mu \nu}_{\mathrm{ext}}({\bf{k}}) = -\frac{\kappa M}{12} \frac{\eta^{\mu \nu}}{{\bf{k}}^2 + m^2_0}. $$

	We are now ready to compute the tree-level scattering of different quantum particles by an external weak higher-order gravitational field. Nevertheless, since these calculations are very extensive, they were  put in Appendix A.

	The outcome of the  experiments analyzed in Appendix A are summarized in Table 1 \footnote{We point out that the constants $\lambda, \lambda_1, \lambda_2, A_0, A_1, A_2, B_0, B_1, B_2, C_0, C_1, C_2, E_0, E_1, E_2,A'_0, A'_1, A'_2, B'_0, B'_1, B'_2, C'_0, C'_1, C'_2, E'_0, E'_1 , E'_2$   in Table 1 are defined in appendix A.}.
	A cursory glance at this table is enough to convince us that  the unpolarized differential cross sections and, of course, the deflection angles, depend on the spin and energy of the scattered particle.

\begin{table}[h!]
	\caption{Unpolarized differential cross sections for the tree-level scattering of different quantum particles by an external weak higher-order gravitational field, where  $ \theta $  is the scattering angle.}
	\renewcommand{\arraystretch}{2.0}
	\setlength{\tabcolsep}{7pt}
	{\begin{center}
			\begin{tabular}{ccc}
				\hline
				\hline                          
				$m$&$s$&$\frac{d\sigma}{d\Omega}$\\ 
				\hline
				$0$&$0$&$\left(\frac{GM}{\sin^{2}\frac{\theta}{2}}\right)^{2}\Big(1+\frac{\sin^{2}\frac{\theta}{2}-3}{3(1+\frac{\lambda_{2}}{4}\csc^{2}\frac{\theta}{2})}-\frac{\sin^{2}\frac{\theta}{2}}{3(1+\frac{\lambda_{0}}{4}\csc^{2}\frac{\theta}{2})}\Big)^{2}$\\
				$\neq0$&$0$&$\left(\frac{GM}{\sin^{2}\frac{\theta}{2}}\right)^{2}\Big(1+\frac{\lambda}{2}+\frac{\sin^{2}\frac{\theta}{2}-(3+2\lambda)}{3(1+\frac{\lambda_{2}}{4}\csc^{2}\frac{\theta}{2})}-\frac{\sin^{2}\frac{\theta}{2}-\frac{\lambda}{2}}{3(1+\frac{\lambda_{0}}{4}\csc^{2}\frac{\theta}{2})}\Big)^{2}$\\
				$0$&$\frac{1}{2}$&$\left(\frac{GM}{\sin^{2}\frac{\theta}{2}}\right)^{2}\cos^{2}\frac{\theta}{2}\Big(1-\frac{1}{1+\frac{\lambda_{2}}{4}\csc^{2}\frac{\theta}{2}}\Big)^{2}$\\
				$\neq0$&$\frac{1}{2}$&$\left(\frac{GM}{\sin^{2}\frac{\theta}{2}}\right)^{2}\sum_{n=0}^2\Big[\frac{E_{n}\lambda^n}{4}+\frac{A_{n}\lambda^n}{9(4+\lambda_{2}\csc^{2}\frac{\theta}{2})^2}+\frac{B_{n}\lambda^n}{9(4+\lambda_{0}\csc^{2}\frac{\theta}{2})^2}+\frac{C_{n}\lambda^n}{9(4+\lambda_{2}\csc^{2}\frac{\theta}{2})(4+\lambda_{0}\csc^{2}\frac{\theta}{2})}\Big]$\\
				$0$&$1$&$\left(\frac{GM}{\sin^{2}\frac{\theta}{2}}\right)^{2}\cos^{4}\frac{\theta}{2}\Big(1-\frac{1}{1+\frac{\lambda_{2}}{4}\csc^{2}\frac{\theta}{2}}\Big)^{2}$\\
				$\neq0$&$1$&$\left(\frac{GM}{\sin^{2}\frac{\theta}{2}}\right)^{2}\sum_{n=0}^2\Big[\frac{E'_{n}\lambda^n}{12}+\frac{A'_{n}\lambda^n}{27(4+\lambda_{2}\csc^{2}\frac{\theta}{2})^2}+\frac{B'_{n}\lambda^n}{27(4+\lambda_{0}\csc^{2}\frac{\theta}{2})^2}+\frac{C'_{n}\lambda^n}{27(4+\lambda_{2}\csc^{2}\frac{\theta}{2})(4+\lambda_{0}\csc^{2}\frac{\theta}{2})}\Big]$\\
				\hline
				\hline
			\end{tabular}
		\end{center}
	}
\end{table}

	Now, bearing in mind that any experiment carried out to test the bending of the quantum particles requires the knowledge of  the gravitational  deflection angle, which, of course, is an extended object, we come to the conclusion that these results can be correctly interpreted 
	as a violation of the  CEP (which is nonlocal) but not of the WEP (which is local).
	
	An important question must be raised now: is it possible to recover the tree-level deflection angles related to general relativity from  Table 1? The answer is affirmative. Indeed, in the 
	$\lambda_2, \lambda_0 \rightarrow \infty$ limit,  Table 1 reduces to  Table 2 displayed below.
	
	It is worthy of  note that the unpolarized differential cross sections exhibited in Table 2, as well as the corresponding  deflection angles, are dependent on the spin; in addition, for the massive particles, the bending depends on the energy as well. 
	
	Why the Einstein gravitational field perceives the spin? Because there is the presence of  a momentum transfer ${\bf k}$ in the scattering responsible for probing the internal structure  (spin) of the particle. Accordingly,  Einstein's geometrical results are recovered in the ${\bf k} \rightarrow 0$; in other words, in the nontrivial limit of small momentum transfer, which corresponds to a nontrivial small angle limit since $|{\bf k}|= 2 |{\bf p}|\sin \frac{\theta}{2}$, the massive (massless) particles behave in the same way, regardless the spin. In fact, if the spin is `switched off', we find from Table 2 that for $m=0$
	
	\begin{eqnarray}
	\frac{d \sigma}{d \Omega} \sim \frac{16G^2 M^2}{\theta^4},
	\end{eqnarray}
	
	\noindent while for $m\neq 0$,

\begin{table}[h!]
	\caption{Tree-level unpolarized differential cross sections for the scattering of different quantum particles by an external weak Einsteinian gravitational field.} 
	
	\vspace{0.1cm}
	\centering
	\renewcommand{\arraystretch}{2.0}
	\setlength{\tabcolsep}{7pt}
	{\begin{center}
			\begin{tabular}{ccc}
				\hline
				\hline  
				$m$&$s$&$\frac{d\sigma}{d\Omega}$\\
				\hline
				$0$&$0$&$\left(\frac{GM}{\sin^{2}\frac{\theta}{2}}\right)^{2}$\\
				$\neq0$&$0$&$\left(\frac{GM}{\sin^{2}\frac{\theta}{2}}\right)^{2}\Big(1+\frac{\lambda}{2}\Big)^{2}$\\
				$0$&$\frac{1}{2}$&$\left(\frac{GM}{\sin^{2}\frac{\theta}{2}}\right)^{2}\cos^{2}\frac{\theta}{2}$\\
				$\neq0$&$\frac{1}{2}$&$\left(\frac{GM}{\sin^{2}\frac{\theta}{2}}\right)^{2}\Big[\cos^{2}\frac{\theta}{2}+\frac{\lambda}{4}\Big(1+\lambda+3\cos^{2}\frac{\theta}{2}\Big)\Big]$\\
				$0$&$1$&$\left(\frac{GM}{\sin^{2}\frac{\theta}{2}}\right)^{2}\cos^{4}\frac{\theta}{2}$\\
				$\neq0$&$1$&$\left(\frac{GM}{\sin^{2}\frac{\theta}{2}}\right)^{2}\Big[\frac{1}{3}+\frac{2}{3}\cos^{4}\frac{\theta}{2}-\frac{\lambda}{3}\Big(1-\frac{3\lambda}{4}-4\cos^{2}\frac{\theta}{2}\Big)\Big]$\\
				
				\hline
				\hline
			\end{tabular}
	\end{center}}
\end{table}	
	
	\begin{eqnarray}
	\frac{d \sigma}{d \Omega} \sim \frac{16G^2 M^2}{\theta^4}\Big( 1 + \frac{\lambda}{2}\Big)^2.
	\end{eqnarray}
	
	These differential cross sections can be related to a classical trajectory with impact parameter $b$ via the relations $bdb \sim- \frac{d \sigma}{d \Omega } \theta d \theta$.
	As a result, we conclude that for $m=0$
	
	\begin{eqnarray}
	\theta \sim \frac{4GM}{b},
	\end{eqnarray}
	
	\noindent and for $m\neq 0$,
	
	\begin{eqnarray}
	\theta \sim \frac{4GM}{b}\Big( 1 + \frac{\lambda}{2} \Big).
	\end{eqnarray}
	
	\noindent The former equation gives the gravitational deflection angle for a massless  particle --- a result foreseen by Einstein a long time ago; whereas the latter just gives the  prediction of general relativity for the bending of a massive particle by an external weak gravitational field \cite{22}. The results of Table 2, in short, reproduce for small angles those predicted by Einstein's geometrical theory, confirming in this way the accuracy of our analytical computations. Note that since $\lambda\equiv \frac{m^2}{{\bf p}^2}= \frac{1 - {\bf v}^2}{{\bf v}^2}$, with ${\bf v}$ being the velocity of the ingoing particle, Eq. (11) tells us that for $|{\bf v}| \ll 1, \; \; \theta \rightarrow \frac{2GM}{b{\bf v}^2}$, which is nothing but Newton's prediction for the gravitational deflection angle; this equation reproduces also Eq. (10) in the $|{\bf v}| \rightarrow 1$ limit. Interestingly enough, since $\lambda  \equiv \frac{m^2}{{\bf p}^2}= \frac{m^2}{E^2 - m^2}$,  for $\frac{m}{E} \ll 1$ Eq. (11) leads  to the result
	
	\begin{eqnarray}
	\theta \sim \frac{4GM}{b} \Big( 1 + \frac{m^2}{2 E^2}\Big),
	\end{eqnarray}
	
	\noindent which was recently utilized to find an upper bound on the photon mass \cite{23,24,25}.

	\section{Violations of the CEP but not of the WEP at the quantum level}
	
	We discuss below two interesting  quantum violations of the CEP but not of he WEP in the Earth gravitational field.
	
	\subsection{COW experiment}
	
	By the mid-1970s, a few years after the publication of  Greenberger's article, using a neutron interferometer, Collela, Overhauser, and Werner \cite{2} analyzed  the quantum mechanical shift of the neutrons caused by the interaction with Earth's gravitational field. Let us then compute the mentioned phase shift.  To accomplish this task, we  make use of a nonintegrable phase shift approach  to gravitation built out  utilizing the similarity of  teleparallel gravity with electromagnetism \cite{26}.
	
	Electromagnetism, as is well known, possesses in addition to the usual differential formalism also a {\it global} formulation in terms of a nonintegrable phase factor \cite{27}. Accordingly,  it can be considered  as the gauge -invariant action  of a nonintegrable (path-dependent) phase factor. As a result, for a particle with electric charge  $e$ traveling from an initial point P to a final point Q, the phase factor assumes the form
	
	\begin{eqnarray}
	\Phi_e(P|Q)= \exp{\Big[ie \int_P^Q{A_\mu dx^\mu\Big]}},
	\end{eqnarray}
		
		\noindent where	 $A_\mu $ is the electromagnetic  gauge potential. Note that the  electromagnetic  phase factor   can also be written as 
		
		\begin{eqnarray}
		\Phi_e (P|Q)= \exp{[iS_e]},
		\end{eqnarray}
		
		\noindent where  $S_e$ is the action integral describing the interaction of the charged particle with the electromagnetic field.
		
		Now, in the teleparallel approach to gravity, the fundamental field describing gravitation is the translational gauge potential $B^a\;_\mu$. Consequently, the action integral concerning the interaction of a particle of mass $m$ with a gravitational field is given by \cite{28} 
		
		\begin{eqnarray}
		S_g= \int_P^Q{m B^a_{~\mu} u_ a dx^\mu}.
		\end{eqnarray}
		\noindent So, the corresponding gravitational nonintegrable phase factor turns out to be
		
		\begin{eqnarray}
		\Phi_g(P|Q)=\exp\Big[ im \int_P^Q {B^a_{~\mu} u_a dx^\mu}\Big].
		\end{eqnarray}
		\noindent It is worthy of mention that similarly to the electromagnetic phase factor, it represents the {\it quantum} mechanical law that replaces the {\it classical } gravitational Lorentz force equation \cite{29}.

		Keeping in mind that  a Newtonian gravitational field is characterized by the condition that only $B^0\;_0   \neq 0$, and taking into account that $u^0= \gamma \simeq1$ for thermal neutrons, the gravitational phase factor becomes 
		
		\begin{eqnarray}
		\Phi_g(P|Q)= \exp\Big[ m \int_P^Q{B_{00}dt}\Big]. 
		\end{eqnarray}
		
		In the Newtonian approximation the above expression reduces to
		
		\begin{eqnarray}
		\Phi_g(P|Q)= \exp\Big[ img  \int_P^Q{z(t) dt}\Big] \equiv \exp\;i \phi,
		\end{eqnarray}
		
		\noindent where  $g$ is the gravitational acceleration and $z$ is the distance from the Earth taken from some reference point.

		We are now ready to calculate the phase $\phi$ through the two trajectories of figure 2, assuming that the segment AC is at $z=0$. For trajectory ACD we promptly obtain

		\begin{figure}[h!]
			\centering
			\includegraphics[scale=0.5]{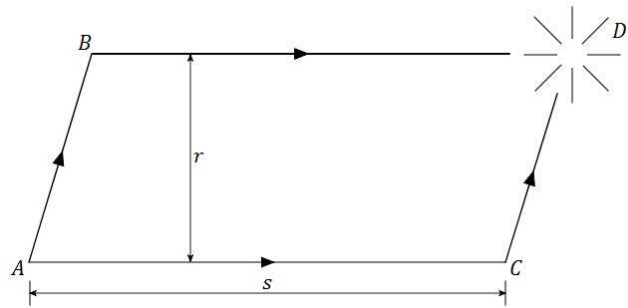}
			\caption{Experiment  to detect gravity-induced quantum interference.}
			\label{fig}
		\end{figure}

		\begin{eqnarray}
		\phi_{\mathrm{ACD}}= mg\int_C^D z(t) dt.
		\end{eqnarray}
		
		Trajectory ABD gives in turn 
		
		\begin{eqnarray} 
		\phi_{\mathrm{ABD}}= mg\int_A^B z(t)dt + mgr\int_B^D dt. 
		\end{eqnarray}
		
		Bearing in mind that the neutron velocity is constant along the segment  BD, we find that
		
		\begin{eqnarray}
		\int_B^D dt \equiv \frac{s}{v}= \frac{sm \lambda}{2\pi}
		\end{eqnarray}
		
		\noindent where $\lambda= \frac{2 \pi}{mv}$ is the de Broglie wavelength  related to the neutron.
		
		Therefore, 
		
		\begin{eqnarray}
		\Delta \phi= \phi_{\mathrm{ABD}}- \phi_{\mathrm{ACD}}=\frac{rs g \lambda m^2}{2\pi}.
		\end{eqnarray} 
		
		So, we come to the conclusion that the phase shift obtained in  COW experiment  is dependent on the neutron mass. This landmark experiment reflects a divergence  between the CEP and quantum mechanics.  Note, however, that  COW phase shift  between the  two neutron paths in   which  these particles are traveling at  different heights in a gravitational field, depends on the (macroscopic) area of the quadrilateral  formed by the neutron paths, being as a consequence a nonlocal effect. Thus,  COW experiment does not violate the WEP.
		
		We call attention to the fact that more  recent and more  accurate experiments have been performed since  COW experiment (1975) in order to  test the WEP on microscopic system via atom interferometry \cite{30,31}. Again, these  experiments  are in accord with the WEP (they are nonlocal) but disagree with the CEP.
		
		\subsection{Free fall in a constant gravitational field  of a massive object described by its wave-function $\Psi$ }
		
		Consider now the  interesting but simple case of free fall in a constant gravitational field
		of a massive object quantum mechanical described by its wave function $\Psi$. We suppose that the wave-function is initially Gaussian.
		
		In this case the Schr\"{o}dinger equation must be satisfied with the Hamiltonian
		
		\begin{eqnarray}
		H= \frac{p^2}{2m} + mgz.
		\end{eqnarray}
		
		The time of flight of the particle at hand can be computed from some initial position $z_0$ up to $z=0$, where the initial position is determined by the expectation value $z_0=\langle \Psi_0|z|\Psi_0 \rangle $ of the position in the Gaussian initial state $\Psi_0$. Now, although the time of flight is statically distributed with the mean value  agreeing  with the classical universal value	
		
		\begin{eqnarray}
		T= \sqrt{\frac{2z_0}{g}},
		\end{eqnarray}
		\noindent the standard deviation of the measured values of the time of flight around $T$ depends on the mass of the particle
		
		\begin{eqnarray}
		\sigma= \frac{2 \pi}{\Delta_0 mg},
		\end{eqnarray}
		\noindent being $\Delta_0$ the width of the initial Gaussian wave packet.
		
		Therefore, we arrive at  the conclusion that in this sense the quantum motion of the particle is non-universal since it depends on the  value of its mass, which of course violates the CEP but not the WEP (the particle is an object extended in space).

		\section{Final remarks}

		Two  semiclassical   examples  that  violate  the CEP but not the WEP in an external gravitational field,  were discussed from a theoretical point of view: Greenberger gravitational Bohr atom and the deflection at the tree level of different quantum particles owed to an external weak higher-order gravitational field. In this latter  case the bending is dependent on the spin and energy of the scattered particle.  We analyzed also an experiment similar to the one just described  where  the external weak higher-order gravitational field is replaced by  an external weak Einsteinian gravitational field which also violates the CEP but not the WEP.

		Two quantum examples that also agree with the WEP but are not in accord with the CEP were analyzed afterwards: COW experiment and free fall in a constant gravitational field of a massive object described by its wave-function $\Psi$.
		
		Now, among the four examples studied in this work only one is based on a experimental test: COW experiment. For this reason we shall elaborate a bit more  on the aforementioned test. Although COW experiment was conducted in 1975, a more accurate  version of the same was performed in 1997 \cite{32}, and its authors reported that in this experiment the gravitationally induced phase shift of the neutron was measured with a statistical uncertainty of order 1 part in 1000 in two different interferometers. A discrepancy between the theoretically predicted and experimentally  measured value of the phase shift due to  gravity was also observed at the 1$\%$ level. Extensions to the theoretical description  of the shape of a neutron interferogram as function of the tilt in a gravitational  field  were discussed and compared with experiment as well. It is worthy of note that past experiments have  verified the quantum-mechanical equivalence of gravitational and inertial masses to a precision of about  1$\%$.
		
		We call attention to the fact that a phase shift of the form given in Eq. (22) would be predicted for a quantum -mechanical particle in the presence of 
		any scalar potential;  in our case is the Newtonian gravitational potential. In order to fully describe this effect we need only quantum mechanics and Newton theory. Therefore,  no metric description of gravity is necessary. This phenomenon, of course, is unexplainable  by classical Newtonian gravity. Undoubtedly, COW experiment represents the first evidence  of gravity interacting in a truly quantum mechanical way. Nevertheless, from the viewpoint of quantum theory, this effect is well understood as a scalar Aharanov-Bohm effect and manifests similarly for electric charges in electric potentials \cite{33,34}.
		
		We point out that the  references  \cite{35, 36, 37,38} may be helpful  for those interested in  investigations   similar in a sense  to those dealt with in the present work.

		Last but not least, we would like to draw the reader's  attention to fact the examples discussed in this article seem to indicate that,  at  first sight, the only possibility of   violating the WEP is through local experiments.

	%	\ack
		The authors are very grateful to FAPERJ and CNPq for their financial support.
		
		\appendix
		\section{Unpolarized differential cross sections for  tree-level scattering of different quantum particles by an external weak higher-order gravitational field}

		\subsection{Spin-0 particles}
		The Lagrangian  for a   massive scalar field minimally coupled to gravity  can be written as 
		
		\begin{eqnarray}
		{\cal {L}} = \frac{\sqrt{-g}}{2}\Big( g^{\mu \nu}\partial_\mu \phi \partial_\nu \phi- m^2 \phi^2 \Big),
		\end{eqnarray}
		\noindent  and  leads to first order in $k$ to the following  Lagrangian for the interaction of a scalar field with a weak gravitational field 
		\begin{eqnarray}
		{\cal{L}}_{\mathrm{int}}= -\frac{\kappa}{2} h^{\mu \nu}\Big[\partial_\mu \phi \partial_\nu  \phi -\frac{1}{2} \eta_{\mu \nu}\Big(\partial_\alpha \phi \partial^\alpha \phi - m^2 \phi^2 \Big) \Big]. \nonumber
		\end{eqnarray}
		
		From the preceding Lagrangian we promptly obtain the vertex for the process depicted in figure A1
		
		\begin{eqnarray}
		V({\bf p}, {\bf q})= -\frac{\kappa}{2} h^{\mu \nu }_{\mathrm{ext}}({\bf{k}})\Big[ p_\mu q_\nu + p_\nu q_\mu -\eta_{\mu \nu}(p \cdot q -m^2) \Big], \nonumber
		\end{eqnarray}
		\noindent where the external field is a weak higher-order gravitational field.
		
		\begin{figure}[h!]
			\centering
			\includegraphics[scale=0.54]{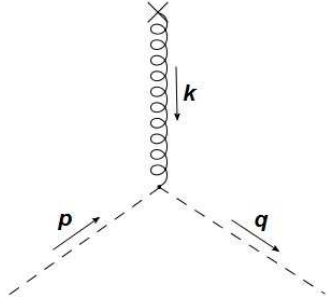}
			\caption{Feynman diagram for the interaction between a  spin-0 particle   and an external  weak  gravitational field.}
			\label{fig}
		\end{figure}

		Now, the differential cross section for the process above reads
		
		\begin{eqnarray} 
		\frac{d \sigma}{d \Omega}= \frac{{|\cal{M}|}^2}{(4 \pi)^2},
		\end{eqnarray}
		\noindent where   the Feynman amplitude ${\cal{M}}$ coincides with $V({\bf p}, {\bf q})$.
		
		Accordingly, the differential cross section for the  tree-level scattering of a massive  spin-0 particle by an external weak higher-derivative gravitational field assumes the form

		\begin{eqnarray}
		\frac{d \sigma}{d \Omega}=\Bigg[ \frac{GM}{\sin^2 \frac{\theta}{2}}\Bigg]^2 \Bigg[1 + \frac{\lambda}{2} + \frac{\sin^2 \frac{\theta}{2} - (3 + 2 \lambda)}{3(1 + \frac{1}{4} \lambda_2 \csc^2 \frac{\theta }{2})} +  \frac{\frac{1}{2}\lambda- \sin^2 \frac{\theta}{2}}{3(1 + \frac{1}{4} \lambda_0 \csc^2 \frac{\theta }{2})} \Bigg]^2,
		\end{eqnarray}
		\noindent where 
		\begin{eqnarray}
		\lambda \equiv \frac{m^2}{{\bf p}^2}, \;\; \lambda_2 \equiv \frac{M^2_2}{{\bf p}^2}, \;\; \lambda_0 \equiv \frac{M^2_0}{{\bf{p}}^2}.
		\end{eqnarray}
		\noindent Now, since  $E^2= {\bf p}^2(1+\lambda)$ where $E$ is the particle energy,  $\lambda_2$ and $\lambda_0$ can be written as
		\begin{eqnarray}
		\lambda_2= \frac{(1+ \lambda)M^2_2}{E^2}, \; \;   \lambda_0= \frac{(1 +\lambda )M^2_0}{E^2},
		\end{eqnarray}
		\noindent  which clearly  shows that  all the parameters in Eq. (A.4) are energy dependent.
		
		In the $m\rightarrow 0 \;$ limit, we get the differential cross section for  tree-level scattering of a massless spin-0 boson by an external weak higher-derivative gravitational field

		\begin{eqnarray}
		\frac{d \sigma}{d \Omega}=\Bigg[ \frac{GM}{\sin^2 \frac{\theta}{2}}\Bigg]^2 \Bigg[1 
		+ \frac{\sin^2 \frac{\theta}{2} - 3 }{3(1 + \frac{1}{4} \lambda_2 \csc^2 \frac{\theta }{2})} -  \frac{ \sin^2 \frac{\theta}{2}}{{3(1 + \frac{1}{4} \lambda_0 \csc^2 \frac{\theta }{2})}} \Bigg]^2.
		\end{eqnarray}
		\noindent Note that if $m=0$, $\lambda_2= \frac{M^2_2}{E^2}\;$ and $ \lambda_0=\frac{M^2_0}{E^2}$. Here we are using the same symbols for denoting the parameters $\lambda_2$ and $\lambda_0$  as those utilized  for the massive case since their meaning are  quite clear from the context. Therefore, from now on these symbols will utilized for both massive and massless particles.
		\subsection{Spin-1/2 particles}
		As is well known, the gravitational  Lagrangian for a  massive fermion is given by \cite{39} 
		
		\begin{eqnarray}
		{\cal{L}}= \sqrt{-g}\Bigg[ \frac{i}{2} \Big(\bar{\psi} \gamma^\mu \overset{\rightarrow}{\nabla}_\mu \psi - \bar{\psi} \overset{\leftarrow}{\nabla}_\mu\gamma
		^\mu \psi \Big) - m \bar{\psi}\psi \Bigg],
		\end{eqnarray}
		\noindent with the notation

		\begin{eqnarray}
		\gamma^\mu= \gamma^p e^\mu_p, \; \overset{\rightarrow}{\nabla}_\mu \psi= \partial_\mu \psi+ iw_\mu \psi, \; \bar{\psi} \overset{\leftarrow}{\nabla}_\mu= \partial_\mu \bar{\psi} - i \bar{\psi}w_\mu. \nonumber
		\end{eqnarray}
		\noindent Here $e^\nu_n \equiv \eta_{m n }g^{\mu \nu}e^m_\mu (x)$ is a different type of vierbein  where the $ m$ index  is lowered with the Minkowski metric $\eta_{n m}$, while the $\mu$ index is  raised with $g^{\mu \nu}$; whereas the  field connection $w_\mu(x)$ is expressed in terms of the tetrads as
		
		\begin{eqnarray}
		w_\mu(x)= \frac{1}{4}\sigma^{m n}\Big[ e^\nu_m \Big(\partial_\mu e_{n\nu} - \partial_\nu e_{n \mu}\Big) + \frac{1}{2}e^\rho_m e^\sigma_n \Big( \partial_\sigma e_{l \rho} -  \partial_\rho e_{l\sigma} \Big)e^l_\mu - (m \leftrightarrow n)\Big], \nonumber
		\end{eqnarray}
		\noindent where the Dirac matrices  are denoted by $\gamma^n$, and $\sigma^{m n}= \frac{i}{2}[\gamma^m, \gamma^n]$.

		Keeping in mind that to order $k$  \cite{40}
		
		\begin{eqnarray}
		e^m_\mu=\delta^m_\mu + \frac{\kappa}{2}h^m_\mu + {\cal{O}}(k^2),
		\end{eqnarray}
		\noindent we find  that  within this approximation the Lagrangian for the interaction of a fermion with a weak gravitational field has the form
		
		\begin{eqnarray}
		{\cal{L}}_{\mathrm{int}}=&& -\frac{\kappa}{2}h_{\mu \nu}\Bigg\{ \frac{i}{2} \Big[ \Big( \bar{\psi} \gamma^\mu \partial^\nu \psi - \partial^\nu \bar{\psi} \gamma^\mu \psi \Big) \nonumber \\ &&- \eta^{\mu \nu} \Big( \bar{\psi} \gamma^\alpha \partial_\alpha \psi - \partial_\alpha \bar{\psi} \gamma^\alpha \psi \Big) \Big] + \eta^{\mu \nu} m \bar{\psi} \psi \Bigg\}.
		\end{eqnarray}
		
		It follows that the vertex for the process shown in figure A2 reads

		\begin{eqnarray}
		V({\bf p}, {\bf q})= \frac{\kappa}{8} h^{\mu \nu }_{\mathrm{ext}}({\bf{k}})\Bigg[ 2\eta_{\mu \nu}\Big(\slashed{p} + \slashed{q} -2m\Big) - \gamma_\mu ( p + q )_\nu  - ( p +q)_\mu \gamma_\nu \Bigg], 
		\end{eqnarray}
		\noindent where $h^{\mu \nu }_{\mathrm{ext}}({\bf k})$ is given by (7).

		\begin{figure}[h!]
			\centering
			\includegraphics[scale=0.54]{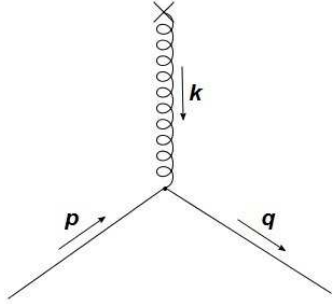}
			\caption{Diagram for the interaction of a  fermion   with an external  weak  gravitational field.}
			\label{fig}
		\end{figure}
		
		The unpolarized differential cross section for the process at hand, in turn, is given by
		
		\begin{eqnarray}
		\frac{d \sigma}{d \Omega}= \frac{(2m)^2}{(4\pi)^2} \frac{1}{2} \sum_{r', r} \Big|{\cal{M}}_{r',r}\Big|^2,
		\end{eqnarray}
		
		\noindent where

		$${\cal{M}}_{r',r}= \bar{u}_{r'}({\bf q})V({\bf p}, {\bf q})u_r {(\bf p)}.$$

		Taking the relation 
		
		\begin{eqnarray}
		\sum_{r=1}^2 u_r({\bf p})\bar{u}_r({\bf p}) =\frac{\slashed{p} + m}{2m} 
		\end{eqnarray}
		\noindent into account, we find that the unpolarized  differential cross section for the scattering of a  massive fermion by an external weak  higher-order gravitational field reads 
		
		\begin{eqnarray}
		\frac{d \sigma}{d \Omega}=&&\Bigg( \frac{GM}{\sin^2 \frac{\theta}{2}}\Bigg)^2  \sum_{n=0}^2 \Bigg[\frac{E_n (\lambda)^n}{4}+ \frac{A_n (\lambda)^n }{9(4 + \lambda_2 \csc^2 \frac{\theta}{2})^2 } \nonumber  \\ &&+\frac{B_n (\lambda)^n}{9(4 + \lambda_0 \csc^2 \frac{\theta}{2})^2} + \frac{C_n (\lambda)^n}{9(4 + \lambda_2 \csc^2 \frac{\theta}{2})(4 + \lambda_0 \csc^2 \frac{\theta}{2})} \Bigg], \nonumber
		\end{eqnarray}
		\noindent where
		$$E_0=4 \cos^2  \frac{\theta}{2}, \;\;E_1=3 \cos^2 \frac{\theta}{2} +1, \;\; E_2=1;$$
		
		$$A_0= -(72\cos^2\frac{\theta}{2})\lambda_2 \csc^2 \frac{\theta }{2} -144\cos^2\frac{\theta}{2},$$

		$$A_1=-(60\cos^2\frac{\theta}{2})\lambda_2\csc^2\frac{\theta}{2} - 24 \lambda_2\csc^2\frac{\theta}{2} - 112 \cos^2 \frac{\theta}{2} -32, $$

		$$A_2= -24 \lambda_2 \csc^2\frac{\theta}{2} - 32; $$

		$$ B_0=0,$$ $$B_1= (6\cos^2\frac{\theta}{2}) \lambda_0  \csc^2 \frac{\theta}{2}+ 6 \lambda_0 \csc^2 \frac{\theta}{2}\ + 20\cos^2\frac{\theta}{2} + 28,$$

		$$ B_2=6 \lambda_0 \csc^2 \frac{\theta}{2} + 28;$$
		
		$$C_0=0, \;\; C_1= -16 \cos^2 \frac{\theta}{2} - 32, \;\; C_2= -32. $$
		
		In the $m \rightarrow 0$ limit, we obtain the differential cross section  for a massless fermion
		
		\begin{eqnarray}
		\frac{d \sigma}{d \Omega}=\Bigg( \frac{GM}{\sin^2 \frac{\theta}{2}}\Bigg)^2 \cos^2\frac{\theta}{2}\Bigg[ 1 - \frac{1}{1+ \frac{1}{4}\lambda_2 \csc^2 \frac{\theta }{2}}\Bigg]^2.
		\end{eqnarray}
		
		\subsection{Spin-1 particles}
		The gravitational Lagrangian for a massive photon  can be written as 
		
		\begin{eqnarray}
		{\cal{L}}= \sqrt{-g}\Bigg[ -\frac{1}{4} g^{\mu \alpha} g^{\nu \beta}F_{\mu \nu} F_{\alpha \beta} + \frac{m^2}{2}g^{\mu \nu} A_\mu A_\nu\Bigg],
		\end{eqnarray}

		\noindent from which we trivially obtain the Lagrangian for the interaction  of a massive photon  with a weak gravitational field
		\begin{eqnarray}
		{\cal{L}_{\mathrm{int}}}= -\frac{\kappa}{2}h^{\mu \nu} \Big[ \frac{1}{4}\eta_{\mu \nu}F^2_{\alpha \beta}- F_\mu^{\;\alpha} F_{\nu \alpha} + m^2 \Big( A_\mu A_\nu - \frac{1}{2}\eta_{\mu \nu}A^2_\alpha \Big) \Big]. \nonumber
		\end{eqnarray}

		Accordingly, the vertex for process represented in figure A3 is given by
		
		\begin{eqnarray}
		V_{\alpha \beta}({\bf p}, {\bf q})=&& -\frac{\kappa}{2} h^{\mu \nu }_{\mathrm{ext}}({\bf{k}}) \Bigg[\Big( \eta_{\alpha \beta} \eta_{\mu \nu } - \eta_{\alpha  \mu} \eta_{\beta \nu} - \eta_{\alpha \nu} \eta_{\beta \mu}\Big) \nonumber \\ && \Big(p \cdot q - m^2 \Big) - \eta_{\alpha \beta}p_\nu q_\mu + \eta_{\mu \beta} p_\nu q_\alpha - \eta_{\mu \nu}p_\beta q_\alpha  \nonumber \\ &&+ \eta_{\alpha \nu}p_\beta q_\mu + \eta_{\beta \nu} p_\alpha q_\mu - \eta_{\mu \nu}p_\alpha q_\beta + \eta_{\alpha \mu }p_\nu q_\beta \Bigg], \nonumber
		\end {eqnarray}
		where the external field is a weak higher-order gravitational fied.
		
		\begin{figure}[h!]
		\centering
		\includegraphics[scale=0.54]{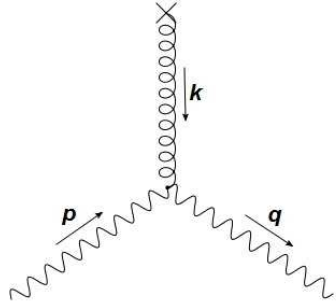}
		\caption{ Diagram for the interaction between a  photon   and an external  weak  gravitational field.}
		\label{fig}
		\end{figure}
		
		Therefore, the unpolarized differential cross section for the process above can be written as
		
		\begin{eqnarray}
		\frac{d \sigma}{d \Omega}= \frac{1}{(4\pi)^2} \frac{1}{3} \sum_{r', r} \Big|{\cal{M}}_{r',r}\Big|^2,
		\end{eqnarray}
		
		\noindent where 
		
		\begin{eqnarray}
		{\cal{M}}_{r',r}= \epsilon^\alpha_{r'}({\bf q}) V_{\alpha \beta}({\bf p}, {\bf q}) \epsilon^\beta_r({\bf p}).
		\end{eqnarray}
		\noindent Here $\epsilon^\beta_r({\bf p})$ and $ \epsilon^\alpha_{r'}({\bf q})$ are respectively the ingoing and outgoing photon polarizations.

		Now, bearing in mind that
		
		\begin{eqnarray}
		\sum_{r=1}^3 \epsilon^\mu_r({\bf p}) \epsilon^\nu_r ({\bf p})= -\eta^{\mu \nu} + \frac{p^\mu  p^\nu}{m^2},
		\end{eqnarray}
		
		\noindent we come to the conclusion that
		
		\begin{eqnarray}
		\frac{d \sigma}{d \Omega}=&&\Bigg( \frac{GM}{\sin^2 \frac{\theta}{2}}\Bigg)^2  \sum_{n=0}^2 \Bigg[\frac{E'_n (\lambda)^n}{12}+ \frac{A'_n (\lambda)^n}{27(4 + \lambda_2
			csc^2 \frac{\theta}{2})^2} \nonumber  \\ &&+\frac{B'_n (\lambda)^n}{27(4 + \lambda_0csc^2 \frac{\theta}{2})^2} + \frac{C'_n (\lambda)^n}{27(4 + \lambda_2 csc^2 \frac{\theta}{2})(4 + \lambda_0 csc^2 \frac{\theta}{2})} \Bigg], \nonumber
		\end{eqnarray}
		\noindent where
		$$E'_0=4  + 8\cos^4  \frac{\theta}{2}, \;E'_1= -4 + 16 \cos^2 \frac{\theta}{2}, \; E'_2=3;$$
		
		\begin{eqnarray}
		A'_0=&& -128 - 32\cos^2 \frac{\theta}{2} - 272 \cos^4\frac{\theta}{2} + 168 \lambda_2 \nonumber \\ &&+ 144 \lambda_2\cos^2 \frac{\theta}{2 }- 216 \lambda_2 \csc^2 \frac{\theta}{2},   \nonumber
		\end{eqnarray}
		
		$$A'_1=160 - 592\cos^2 \frac{\theta}{2 }+ 324\lambda_2 - 252 \lambda_2 \csc^2 \frac{\theta}{2},$$

		$$A'_2= -96 - 72\lambda_2 \csc^2 \frac{\theta}{2}; $$

		$$ B'_0=-80 + 64\cos^2 \frac{\theta}{2 }+ 16 \cos^4 \frac{\theta}{2} -24 \lambda_0,$$

		$$B'_1= 16 + 128 \cos^2\frac{\theta}{2} - 36\lambda_0 + 36 \lambda_0 \csc^2\frac{\theta}{2},$$

		$$ B'_2= 84 + 18 \lambda_0 csc^2 \frac{\theta}{2};$$
		
		$$C'_0= 64 - 32 \cos^2\frac{\theta}{2} - 32 \cos^4 \frac\theta{2},$$  
		
		$$C'_1= -32 - 112 \cos^2 \frac{\theta}{2},  \;\; C'_2= -96. $$
		
		On the other hand, the gravitational Lagrangian for a  massless photon has the form
		
		\begin{eqnarray}
		{\cal L}= \sqrt{-g}\Big( \frac{1}{4}g^{\mu \alpha} g^{\nu \beta}F_{\mu \nu}F_{\alpha \beta}\Big), 
		\end{eqnarray}
		
		\noindent and from it we find the  Lagrangian for the interaction between a massless photon and a weak gravitational 
		
		\begin{eqnarray}
		{\cal{L}}_{\mathrm{int}}= -\frac{\kappa}{2}h^{\mu \nu}\Bigg(\frac{1}{4}\eta_{\mu \nu}F^2_{\alpha \beta} - F^{\; \alpha}_\mu F_{\nu \alpha} \Bigg).
		\end{eqnarray}
		
		It follows then that the vertex  for the interaction  of a massless photon with an external weak higher-order gravitational field reads

		\begin{eqnarray}
		V_{\alpha \beta}({\bf p}, {\bf q})=&& -\frac{\kappa}{2} h^{\mu \nu }_{\mathrm{ext}}({\bf{k}}) \Bigg[\Big( \eta_{\alpha \beta} \eta_{\mu \nu } - \eta_{\alpha  \mu} \eta_{\beta \nu} - \eta_{\alpha \nu} \eta_{\beta \mu}\Big)p \cdot q  \nonumber \\ &&- \eta_{\alpha \beta}p_\nu q_\mu \nonumber + \eta_{\mu \beta} p_\nu q_\alpha - \eta_{\mu \nu}p_\beta q_\alpha  + \eta_{\alpha \nu}p_\beta q_\mu  \nonumber \\ &&+ \eta_{\beta \nu} p_\alpha q_\mu - \eta_{\mu \nu}p_\alpha q_\beta + \eta_{\alpha \mu }p_\nu q_\beta \Bigg]. \nonumber
		\end{eqnarray}
		
		Now, the differential cross section for the process under discussion can be written as

		\begin{eqnarray}
		\frac{d \sigma}{d \Omega}= \frac{1}{(4\pi)^2} \frac{1}{2} \sum_{r', r} \Big|{\cal{M}}_{r',r}\Big|^2,
		\end{eqnarray}
		
		\noindent where 
		
		\begin{eqnarray}
		{\cal{M}}_{r',r}= \epsilon^\alpha_{r'}({\bf q}) V_{\alpha \beta}({\bf p}, {\bf q}) \epsilon^\beta_r({\bf p}).
		\end{eqnarray}
		
		Keeping in mind that 
		\begin{eqnarray}
		\sum_{r=1}^{2}\epsilon^\mu_r({\bf p }) \epsilon^\nu_r ({\bf p})=- \eta_{\mu \nu} -\frac{1}{(p \cdot n)^2}\Big[  p^\mu p^\nu - p \cdot n \Big(p^\mu n^\nu + p^\nu n^\mu\Big)  \Big], \nonumber
		\end{eqnarray}
		
		\noindent where $n^2=1$, we arrive at the conclusion that

		\begin{eqnarray}
		\frac{d \sigma}{d \Omega}=\Bigg( \frac{GM}{\sin^2 \frac{\theta}{2}}\Bigg)^2  \cos^4 \frac{\theta}{2} \Bigg( 1- \frac{1}{1 + \frac{1}{4}\lambda_2 \csc^2\frac{\theta}{2}} \Bigg)^2.\nonumber  
		\end{eqnarray}

	\end{document}